\documentclass{iopart}
\usepackage{graphicx}
\usepackage{xcolor}
\begin{document}

\submitto{\NJP}

\title[Transport, Hall Effect, and Collective Dynamics...in Active Chiral Media]{Transport, Hall Effect, and Collective Dynamics of Driven Particles in Active Chiral Media}
\author{C. Reichhardt and C. J. O. Reichhardt$^*$} 
\address{
Theoretical Division and Center for Nonlinear Studies,
Los Alamos National Laboratory, Los Alamos, New Mexico 87545, USA
}
\address{$^*$ Corresponding author}

\ead{cjrx@lanl.gov}

\date{\today}

\begin{abstract}
We examine the dc driven transport of particles with Yukawa interactions through an assembly of chiral active particles undergoing circular motion. The driven particles exhibit a jammed phase, a nonlinear disordered or plastic regime where there is a finite Hall effect produced by interactions with the active chiral particles, and a high drive linear regime where the driven and nondriven particles phase separate and the Hall effect nearly disappears. The chiral particles experience strong drag and Hall effects from the driven particles, and these effects vary monotonically with changing thermal fluctuations or chiral motion parameters. The different dynamical phases are correlated with features in the longitudinal and transverse transport curves, and we map out the phases as a function of active orbit size, thermal fluctuations, and particle density.
\end{abstract}


\maketitle

\section{Introduction}
A variety of systems can be effectively described 
as collections of interacting particles driven over a disordered or
periodic substrate 
\cite{Fisher98,Reichhardt17}.
These include
vortices in type-II superconductors \cite{Blatter94},
magnetic skyrmions \cite{Nagaosa13},
Wigner crystals \cite{Reichhardt01},
colloidal particles \cite{Pertsinidis08,Bohlein12},
emulsions \cite{LeBlay20}, and
active matter assemblies \cite{Morin17,Sandor17a}.
Typically such systems exhibit a pinned state at small drives,
a disordered or fluid flow phase at intermediate drives, and
partially or fully ordered states at higher drives \cite{Reichhardt17}.
The depinning transition and the changes
between different moving states are associated with
features or nonlinearities in the velocity-force curves,
differential mobility curves, and noise fluctuations
\cite{Fisher98,Reichhardt17}.

In active microrheology,
a small number of individually driven particles pass through a
bath of nondriven particles and can
exhibit a depinning threshold as well as
have different types of moving phases
\cite{Hastings03,Habdas04,Squires05,Gazuz09,Gnann11,Wilson11a,Gruber16,Zia18,Yu20}.
Typical active microrheology measurements include
velocity-force relations and thresholds to motion
\cite{Hastings03,Habdas04,Voigtmann13,Benichou13a,Gruber20}.
There can be
changes in the effective viscosity or damping as a function of drive or when the bulk system exhibits a transition from fluid to crystal to glass 
\cite{Habdas04,Zia18} or from jammed to unjammed
\cite{Hastings03,Reichhardt10,Senbil19}.
Although most studies of active microrheology are for a
single driven particle, it has been shown that when multiple (two or more)
particles are driven and move in close proximity to each other, their
drift velocity is higher than that of an isolated driven particle
\cite{Dzubiella03,Reichhardt06,MejiaMonasterio11}.
Most active rheology work
has been done in systems where the
bulk particles are passive or undergo only thermal fluctuations.
A limited amount of studies have been performed on
active rheology in active matter systems \cite{Reichhardt15,Knezevic21,Peng22}.

There are a growing number of realizations of active matter
particle-based systems where chiral dynamics arise,
such as when the background particles are spinning or
moving in circles
\cite{Lemelle10,Kummel13,Nguyen14,Nourhani15,Lowen16,vanZuiden16,Banerjee17,Han17,Ai18,Soni19,Reichhardt19,Lei19,MassanaCid21,Liebchen22,Tan22,Fruchart23,Caprini26}. 
An interesting feature of the active chiral systems is
that they exhibit odd viscosity in which forces on the particles produce
transverse motion, which can give rise to edge currents and a Hall effect
\cite{Banerjee17,Soni19,Han21,Reichhardt22b,Fruchart23,Huang25}. 
In order to apply
active microrheology to chiral systems,
Reichhardt and Reichhardt \cite{Reichhardt19b}
considered a single driven probe particle moving through an assembly 
of nondriven rotating particles.
In addition to
the nonlinear velocity-force curve in the direction of drive, there was
also a nonzero Hall angle that depended nonmonotonically on the drive,
reaching a maximum in the nonlinear regime and
going to zero at large drives.
A finite Hall effect was also observed in
a numerical study of a rotating disk moving through an array of
fixed obstacles \cite{Lou22}.
In both of these cases,
the rotation of either the background particles or the
driven particles generates a skew scattering
that imparts a motion transverse to the driving direction
on the moving particle.
Driven spinning colloids in viscoelastic fluids were also shown to
exhibit a Hall effect \cite{Cao23}.
Other studies have
focused on odd viscosity and Hall effects
for driven particles in chiral systems \cite{Kalz26}
or odd dynamics of passive obstacles in chiral active baths
\cite{Poggioli23,Hargus25}.
A related particle-like system is magnetic skyrmions,
which have a Magnus component to their dynamics
that produces a Hall effect
due to skew scattering off of pinning sites
\cite{Jiang17,Litzius17,Reichhardt22a}.
In studies of driven particles moving through skyrmion arrays,
there is both
a nonlinear Hall effect as well as a boost effect
in which the probe particle
moves faster than the drive force when the Magnus
component converts some of the driving force into
longitudinal motion \cite{Reichhardt21}.

Here we consider the collective dynamics of particles with
Yukawa or screened Coulomb interactions
driven through a chiral bath
of other Yukawa particles that are undergoing circular motion.
We measure the velocity components in the
directions longitudinal and transverse to the drive
for both particle species.
The chiral particles move in a circular orbit with
an orbit radius of $R_a$ at a frequency $\omega$,
while the driven particles experience
an increasing dc drive applied along
a fixed direction.
For small chiral motion, 
at low drives there is a pinned or jammed phase,
while at larger drives the driven particles slip through
the chiral particle lattice and a mixed or fluid-like phase
appears.
When the chiral orbits are larger, the system can form a fluid-like
state even at zero drive 
and there is no longer a jammed phase.
In the disordered or plastic flow regime, the driven particles exhibit
a Hall effect due to the interactions with the chiral particles.
The driven particles also induce a strong 
drag effect on the chiral particles that causes the chiral particles to
move in the direction of the dc drive.
The dragged particles
also show a Hall effect that is
in the opposite direction to
that of the driven particles.
At higher drives, the system 
dynamically phase separates into a laned state
with a strong reduction in the interaction between the two particle species.
Here the transport becomes linear
and the Hall effect on both species goes to zero. 
We show that the drag
effect can be enhanced by
thermal fluctuations in the disordered regime or
by increasing the size of the chiral orbits.
We also discuss how this work is distinct
from earlier work on active rheology in chiral systems,
which focused on disks with short range interactions. 

\section{Simulation}

We simulate a two-dimensional system
of size $L \times L$ with $L=48$ and periodic boundary conditions in the
$x$ and $y$ directions
containing $N$ particles with Yukawa or screened Coulomb interactions
that are undergoing circular motion.
The total particle density is $\rho=N/L^2$.
For a subset of $N_d$ of the particles,
we switch off the circular motion and instead
apply a fixed direction dc drive that we slowly increase from zero to
a maximum drive level.
The ratio of the driven to non driven
particles is $N_d/N$, which we fix to
$N_d/N=0.25$ so that a quarter of the
particles are driven. The number of chiral particles is then $N_c=N-N_d$.
All particles, whether driven or chiral,
interact through a repulsive Yukawa potential
that models charged colloids, given by
$V(R_{ij}) = C e^{-R_{ij}}/R_{ij}$, where $C$ is a prefactor, ${\bf R}_{i(j)}$ is the position of particle $i(j)$, and $R_{ij}=|{\bf R}_i-{\bf R}_j|$.
The particles are initially placed in a triangular lattice,
and we select the driven particles from this lattice
either in a periodic pattern or randomly.
The motion of the particles is overdamped and is obtained
by integrating the following equation of motion:
\begin{equation}
\eta \frac{d {\bf R}_i}{dt} = -\sum_{j\neq i}^{N}\nabla V(R_{ij}) +
{\bf F}_{D}^i + {\bf F}^{T} 
\end{equation}
where the damping coefficient is $\eta = 1.0$.
For the driven particles, which we term species A,
the driving is a dc force of magnitude $F_D$ applied
in the $x$-direction,
${\bf F}^{i}_{D}=F_{D}\hat{\bf x}$.
The active chiral particles,
which we call species B, are subjected to
a rotating drive 
${\bf F}^{i}_D(t)=A\sin(\omega t)\hat{\bf y}+A\cos(\omega t)\hat{\bf x}$
where $t$ is time.
This drive produces a counterclockwise orbit of the active particles,
and the orbit radius increases with increasing drive amplitude $A$ or
with decreasing drive frequency $\omega$.
We measure the velocities averaged over $1\times 10^4$
simulation time steps in the $x$ and $y$ directions,
$\langle V^{A}_{x}\rangle = (2/N_d)\sum^{N_d}_{i=1}\delta(\sigma_{i}-1)({\bf v}_{i}\cdot\hat{\bf x})$,
$\langle V^{A}_{y}\rangle = (2/N_d)\sum^{N_d}_{i=1}\delta(\sigma_{i}-1)({\bf v}_{i}\cdot\hat{\bf y})$,
$\langle V^{B}_{x}\rangle = (2/N_c)\sum^{N_c}_{i=1}\delta(\sigma_{i})({\bf v}_{i}\cdot\hat{\bf x})$,
and $\langle V^{B}_{y}\rangle = (2/N_c)\sum^{N_c}_{i=1}\delta(\sigma_{i})({\bf v}_{i}\cdot\hat{\bf y})$,
where ${\bf v}_{i}$ is the velocity of
particle $i$ and $\sigma_{i}=1$ (0)
indicates that particle $i$ belongs to species A (B).

We consider the effect of finite temperature or thermal fluctuations
${\bf F}^T$ which are modeled as random Langevin kicks
with the following properties:
$\langle F^{T}(t)\rangle = 0$ and
$\langle F^T_i(t)F^T_j(t^\prime)\rangle = 2\eta k_BT\delta_{ij}\delta(t- t^\prime)$.
For a nondriven or nonchiral system at a given density,
there is a well-defined melting temperature $T_m$
which is characterized by the onset of long-time diffusion
and the proliferation of non-six-fold coordinated particles.

We also consider an alternative pair interaction potential
for the particles in the form of a modified Bessel function,
$V(R_{ij}) = F_0K_{0}(R_{ij})$, where $F_0$ is a constant.
This function decays exponentially at large distances and
describes vortex interactions in type-II superconductors as well as certain
screened-charge systems \cite{Reichhardt17}. 

\section{Results}

\begin{figure}
\centering
\includegraphics[width=3.5in]{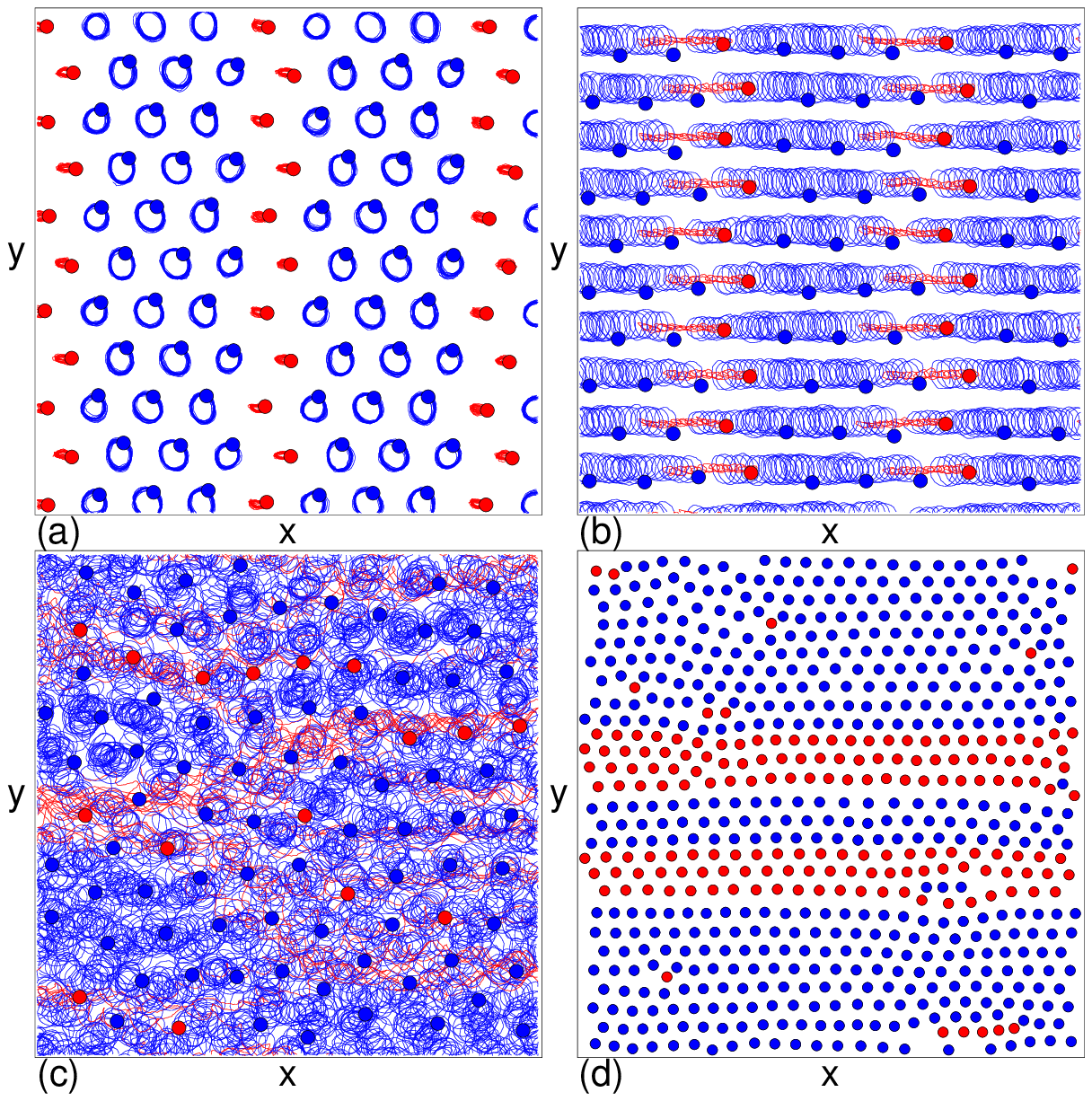}
\caption{Particle locations (circles) and trajectories (lines) in
a portion of the sample for
$\rho=0.29$,
ac drive amplitude $A=2.0$ and frequency $\omega=0.02$,
and temperature $T/T_m=0.2$.
Red indicates the driven species A particles which experience a dc drive
of magnitude $F_D$ in the $x$ direction.
Blue indicates the active chiral species B particles which are subjected
to circular driving.
(a) $F_{D} = 0.0$. (b) The jammed state at $F_{D} = 0.5$.
(c) The disordered flow phase at $F_{D} = 2.5$.
(d) Particle positions only shown for the entire sample
in the high drive phase-separated state at $F_{D} = 5.5$.}
\label{fig:1}
\end{figure}

We first study
system with density
$\rho = 0.29$ and temperature $T/T_m=0.2$, where $T_m$ is the melting
temperature of a nonactive and nondriven system at the same density, and
the active species B particles experience a circular ac drive
with amplitude $A=2.0$ and frequency $\omega=0.02$.
For these parameters,
the system forms a crystal at $F_{D} = 0.0$,
as shown in Fig.~\ref{fig:1}(a), where
we highlight the particle trajectories over a fixed time
to illustrate that the circular orbits of the active chiral
species B particles have
some randomness due to the thermal fluctuations.
The non-active particle species A executes
smaller elliptical orbits due to its coupling
with the rotating species B.
When a finite but small $F_D$ is applied to species A,
all of the particles move as a rigid solid so that species A and species
B have the same velocity. We designate this a jammed phase, and in
Fig.~1(b) we illustrate
the jammed state at $F_{D} = 0.5$, where all particles
show a net translation along $+x$ and there are no plastic deformations.
At higher drives, the system breaks up into a disordered or mixed flow state,
shown in Fig.~\ref{fig:1}(c) at $F_{D} = 2.5$,
where the triangular ordering is lost and
the particles can exchange neighbors.
In the disordered phase, species A
particles are able to move past species B particles, but the species B
particles experience a strong drag effect in the $+x$ direction.
At higher drives, the system forms phase separated lane states of the
type shown in Fig.~\ref{fig:1}(d) at $F_D=5.5$, where we do not plot the
particle trajectories for clarity.
In the laned state, the drag effect from species A on species B
is strongly reduced since the amount of direct interaction between the
two species is minimized.
The drive at which the system transitions
from the jammed phase to the disordered phase
depends on the initial placement of the driven particles.
We used a periodic placement in Fig.~\ref{fig:1}, but if we instead
place the driven particles randomly, the transition to the disordered phase
drops to lower but still finite values of $F_D$, while the drive at
which the system switches from
the disordered to the phase-separated state is unchanged.

\begin{figure}
  \centering
\includegraphics[width=3.5in]{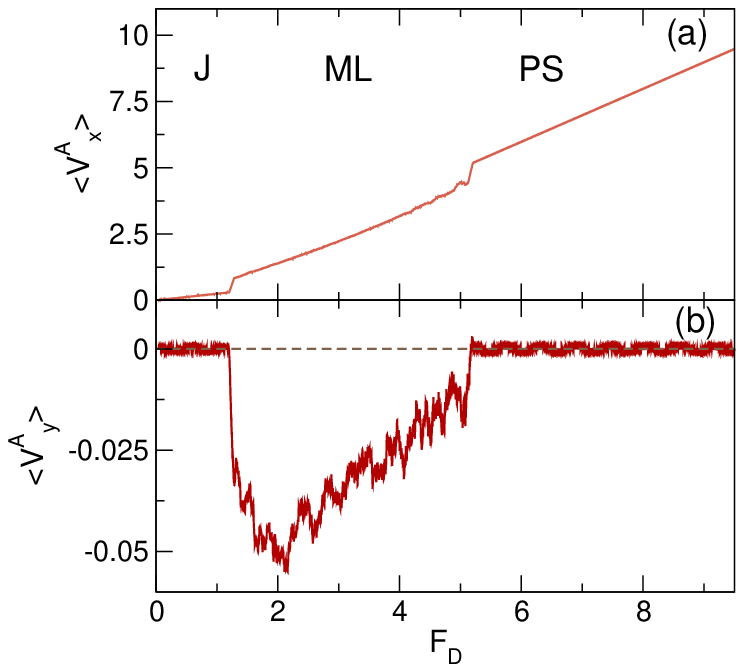}
\caption {
(a) $\langle V^{A}_{x}\rangle$ vs $F_{D}$ for the system from Fig.~\ref{fig:1}
with
$\rho=0.29$, $T/T_m=0.2$, $A=2.0$, and $\omega=0.02$. 
We highlight the jammed (J) phase, disordered or moving liquid (ML) phase,
and phase separated lane state (PS).
(b) The corresponding $\langle V^{A}_{y}\rangle$ vs $F_{D}$
showing that in the disordered flow or ML phase,
there is a finite transverse velocity leading to a finite Hall angle.
}
\label{fig:2}
\end{figure}

In Fig.~\ref{fig:2} we plot $\langle V^{A}_{x}\rangle$
and $\langle V^{A}_{y}\rangle$ versus $F_{D}$
for the system from Fig.~\ref{fig:1}.
For $0 < F_{D} < 1.25$, the system is in the jammed phase, where
$\langle V^{A}_{x}\rangle$ has a low value and
$\langle V^{A}_{y}\rangle = 0.0$.
At the transition from the jammed to the disordered flow phase, there
is a jump
in $\langle V^{A}_{x}\rangle$
accompanied by the onset of a finite and negative value of
$\langle V^{A}_{y}\rangle$.
This indicates that in the disordered flow phase,
the species A particles are now also moving in the $-y$
direction, transverse to the applied driving force $F_D$,
so that species A is exhibiting a finite Hall angle.
The system remains in the disordered flow regime
for $1.3 < F_{D} < 5.0$.
At $F_{D} = 5.2$, $\langle V^A_x\rangle$ jumps up again before
increasing linearly with increasing $F_{D}$.
This second jump in $\langle V^A_x\rangle$
marks the transition from the disordered flow state to the
phase separated lane state.
The transverse velocity
$\langle V^{A}_{y}\rangle$ drops to zero
in the laned state,
indicating the loss of the Hall angle.
The finite Hall angle in the disordered flow phase
appears when the driven particles frequently collide
with the chiral particles and preferentially
scatter in the $-y$ direction.
Close to the transition from the jammed to the disordered flow state, the
Hall angle reaches its maximum magnitude of
$-2.5^\circ$. The Hall angle decreases in magnitude with increasing drive
throughout the disordered flow state before dropping to zero
in the laned state.
The loss of the Hall angle in the laned state occurs because the formation
of lanes minimizes the number of
direct collisions that occur between the driven particles
and the chiral particles, thereby also minimizing the amount
of $y$ direction scattering that occurs.

\begin{figure}
  \centering
\includegraphics[width=3.5in]{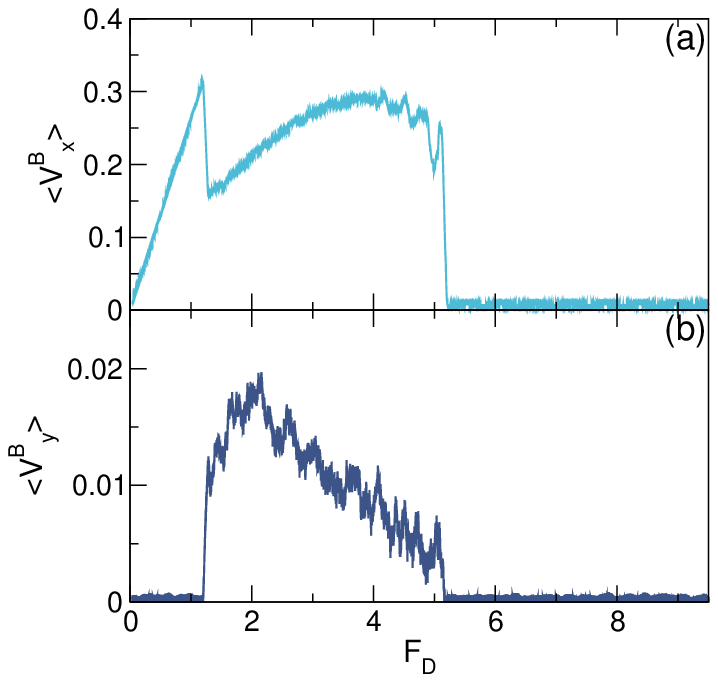}
\caption {$\langle V^{B}_{x}\rangle$ vs $F_{D}$ for the system
from Fig.~\ref{fig:1} with
$\rho=0.29$, $T/T_m=0.2$,
$A=2.0$, and $\omega=0.02$.
(b) The corresponding $\langle V^{B}_{y}\rangle$ vs $F_{D}$
showing that in the disordered flow phase,
there is a finite transverse velocity and
that the Hall effect is in the opposite direction
to that of species A.
}
\label{fig:3}
\end{figure}

In Fig.~\ref{fig:3} we show $\langle V^{B}_{x}\rangle$ and
$\langle V^{B}_{y}\rangle$ versus $F_{D}$
for the same system as in Fig.~\ref{fig:2}.
In the jammed phase,
$\langle V^{B}_{x}\rangle$ increases linearly
because species A and species B are moving together at the same velocity.
In the disordered phase, $\langle V^{B}_{x}\rangle$
initially drops with increasing $F_D$
because the drag effect is reduced when the
driven particles become able
to move past the chiral particles.
There is, however, still a fairly strong drag effect, so that
species B has a finite $x$-direction velocity in the disordered regime.
In the phase separated lane state, $\langle V^{B}_{x}\rangle$
drops nearly to zero
when the direct collisions with the driven particles
and hence the drag are reduced.
The drag velocity of species B is
non-monotonic and increases
to a maximum near $F_{D} = 4.0$
before dropping abruptly at the
transition from the disordered state to the phase separated lane state.
We also find that $\langle V^B_y\rangle$ is
finite and positive in the disordered regime,
and goes to zero in the phase separated lane regime.
This indicates that in the disordered regime, 
the Hall effect on species B is in the positive
$y$ direction and thus has the opposite sign
as the Hall effect of species A.

\begin{figure}
\centering
\includegraphics[width=3.5in]{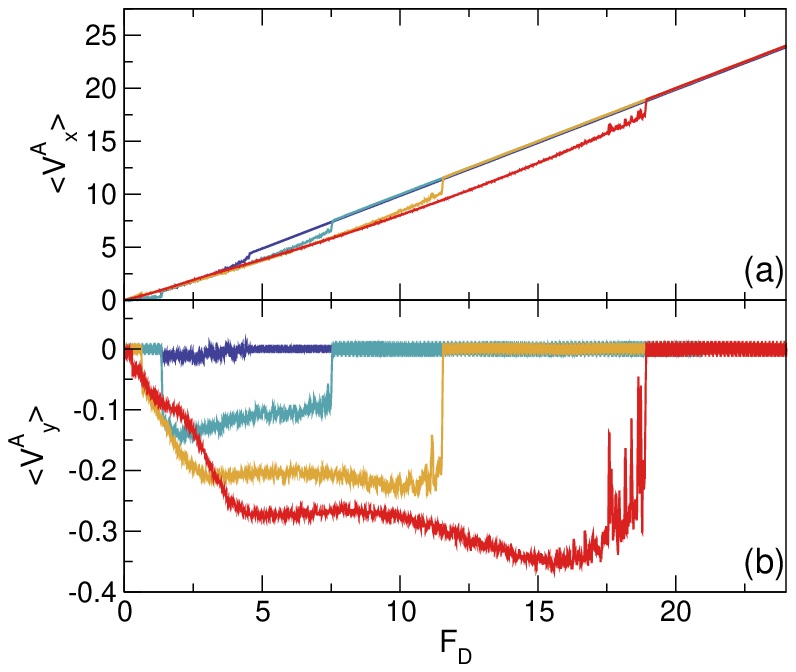}
\caption {(a) $\langle V^{A}_x\rangle$ vs $F_{D}$ for a system with
$\rho=0.29$, $T/T_m=0.2$, and $\omega=0.02$ at
ac drive amplitudes of $A=1.0$ (dark blue),
$4.0$ (light blue),
$6.0$ (orange),
and $8.0$ (red).
(b) The corresponding $\langle V^{A}_y\rangle$ vs $F_{D}$.
The extent of the disordered phase grows with increasing $A$
and the maximum value of $|\langle V^{A}_y\rangle|$ also increases.}
\label{fig:4}
\end{figure}

The extent of the jammed and disordered phases,
the size of the drag velocity,
and the maximum transverse velocities of species A and B
strongly depend on the ac driving amplitude of the chiral particles.
In Fig.~\ref{fig:4}(a) we plot
$\langle V^{A}_x\rangle$ versus $F_{D}$
for the system from Fig.~\ref{fig:2}
at $A = 1.0$, 4.0, 6.0, and $8.0$.
Figure~\ref{fig:4}(b) shows the corresponding
$\langle V^{A}_y\rangle$ versus $F_{D}$ curves.
The jammed phase is absent for $A > 4.0$
because the system is in a fluid state even
at $F_{D} = 0.0$
since the orbits of species B are strongly overlapping,
resulting in unavoidable
collisions with species A.
The transition from the disordered to the phase separated lane state
is marked by
a jump up in $\langle V^{A}_{x}\rangle$,
which shifts to higher values of $F_{D}$ as the ac driving amplitude
increases.
The jammed to disordered and disordered to phase separated lane transitions
are also clearly visible in the $\langle V^A_y\rangle$ curves,
and we find that
the maximum value of
$|\langle V^{A}_y\rangle|$ increases with increasing $A$.

\begin{figure}
\centering
\includegraphics[width=3.5in]{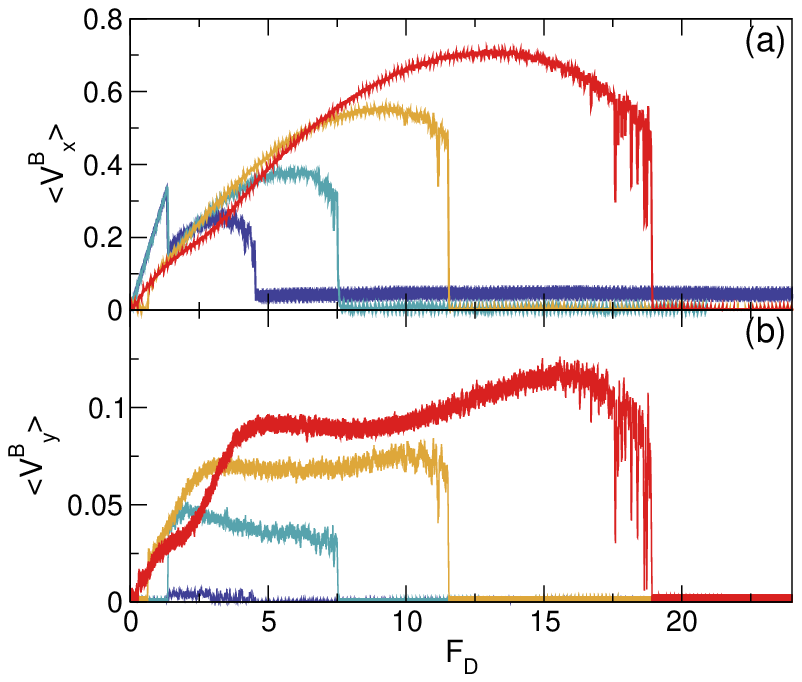}
\caption {(a) $\langle V^{B}_x\rangle$ vs $F_{D}$ for the system from
Fig.~\ref{fig:4} with
$\rho=0.29$, $T/T_m=0.2$,
and $\omega=0.02$ at ac drive amplitudes of
$A=1.0$ (dark blue),
$4.0$ (light blue),
$6.0$ (orange),
and $8.0$ (red).
(b) The corresponding $\langle V^{B}_y\rangle$ vs $F_{D}$.  
}
\label{fig:5}
\end{figure}

In Fig.~\ref{fig:5},
we show $\langle V^B_x\rangle$ and $\langle V^B_y\rangle$ versus $F_D$ at
different ac drive amplitudes
for the same system from Fig.~\ref{fig:4}.
As the ac drive amplitude increases,
the drag effect on species B strongly increases.
The local maximum in $\langle V^B_x\rangle$ shifts to
higher values of $F_{D}$ with increasing $A$
before dropping to a value close to zero in the
phase separated lane state.
The maximum value of $\langle V^{B}_y\rangle$ also increases
with increasing ac drive amplitude,
indicating that the chiral particles are traveling
along the positive $y$-direction.

\begin{figure}
\centering
\includegraphics[width=3.5in]{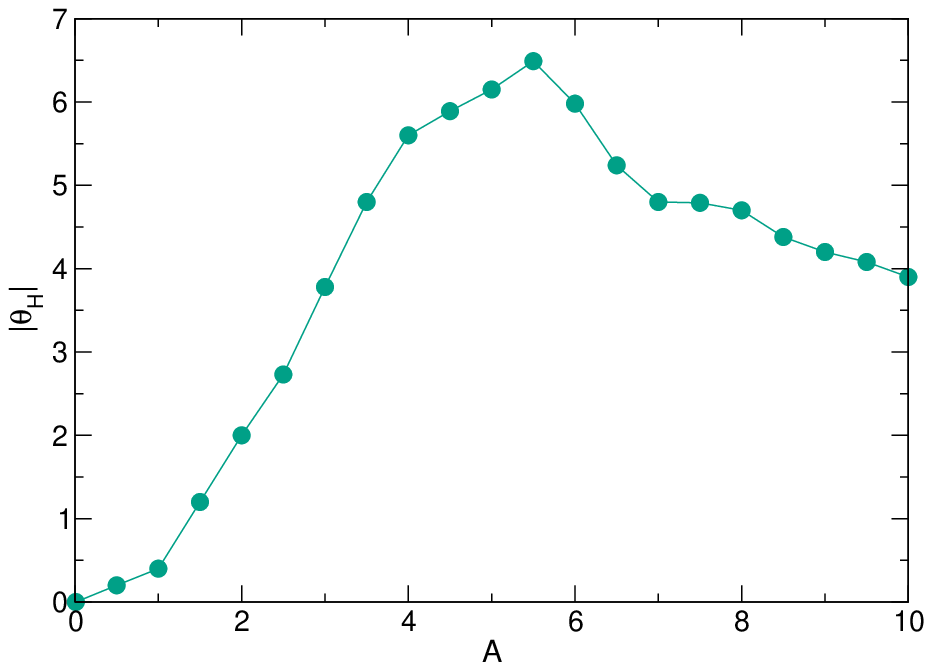}
\caption {The maximum value of the Hall angle $|\theta_H|$
for the driven species A vs ac drive amplitude $A$ from
the system in Fig.~\ref{fig:4} with
$\rho=0.29$,
$T/T_m=0.2$, and $\omega=0.02$.
The overall maximum Hall angle magnitude is
$|\theta_H|=6.5^\circ$.
	}
\label{fig:6}
\end{figure}

In Fig.~\ref{fig:6} we plot the maximum absolute Hall angle
$|\theta_H|$ versus ac drive amplitude for species  A.
In this case the maximum Hall angle does not necessarily
correspond to the maximum value of $|\langle V^A_{y}\rangle|$.
Instead it is given by the maximum ratio
$|\theta_H|=|\langle V^A_{y}\rangle|/|\langle V^A_{x}\rangle|$,
which falls at different values of $F_{D}$ for different $A$.
The overall maximum Hall angle magnitude is close to
$|\theta_H|=6.5^\circ$.

\begin{figure}
\centering
\includegraphics[width=3.5in]{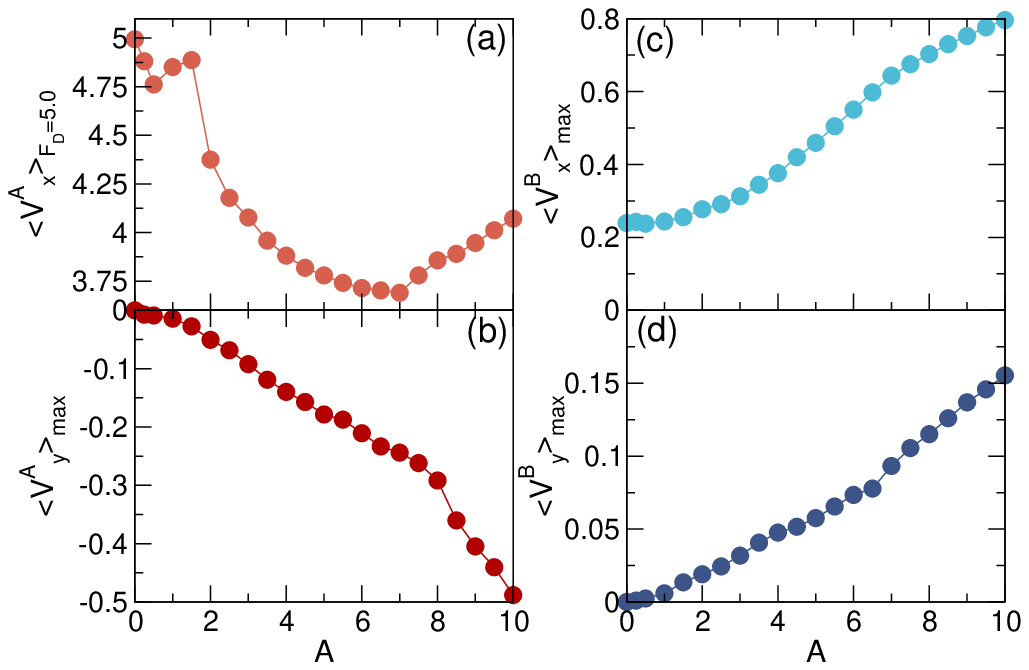}
\caption {Plots for the system from Fig.~\ref{fig:4} with
$\rho=0.29$, $T/T_m=0.2$, and $\omega=0.02$.
(a) The value of
$\langle V^{A}_x\rangle$ at $F_D=5.0$,
$\langle V^A_x\rangle_{F_D=5.0}$,
vs ac amplitude $A$ for the system
from Fig.~\ref{fig:4} at $F_{D} = 5.0$.
(b) The maximum value of
$\langle V^{A}_y\rangle$ in the disordered regime,
$\langle V^{A}_y\rangle_{\rm max}$,
vs $A$.
(c) The maximum value of
$\langle V^{B}_x\rangle$ in the disordered regime,
$\langle V^{B}_x\rangle_{\rm max}$,
vs $A$.
(d) The maximum value of
$\langle V^B_y\rangle$ in the disordered regime,
$\langle V^B_y\rangle_{\rm max}$, vs $A$.
}
\label{fig:7}
\end{figure}

To better characterize the evolution of the drag and transverse velocities
for the system from Fig.~\ref{fig:4},
in Fig.~\ref{fig:7}(a) we
plot the value of $\langle V^A_x\rangle$ in the disordered phase at
$F_D=5.0$, $\langle V^A_x\rangle_{F_D=5.0}$, versus ac drive amplitude $A$.
We use a fixed value of $F_D$ here because $\langle V^A_x\rangle$ never
reaches a maximum but instead increases across the entire range
of $F_D$.
At $F_D=5.0$, the system forms a laned state for
$A < 2.5$ and a disordered state for $A \geq 2.5$.
The drop in $\langle V^{A}_x\rangle_{F_D=5.0}$
at an ac amplitude of $A=2.5$ coincides with the system entering
the disordered state.
As $A$ increases further in the disordered phase,
$\langle V^{A}_x\rangle_{F_D=5.0}$ continues to drop,
indicating an increase in the drag effects.
There is a local minimum in $\langle V^A_x\rangle_{F_D=5.0}$ near
$A=7.0$, after which $\langle V^A_x\rangle_{F_D=5.0}$ increases with
increasing $A$ again.
In Fig.~\ref{fig:7}(b) we plot the maximum value
of $\langle V^{A}_y\rangle$ in the disordered regime,
$\langle V^A_y\rangle_{\rm max}$, versus $A$.
This maximum value falls at
higher values of $F_{D}$
as the ac drive amplitude increases.
When the ac drive amplitude is zero,
we find $\langle V^{A}_y\rangle_{\rm max} = 0.0$ since the species B particles
provide no chiral component to the system.
As $A$ increases, $|\langle V^{A}_y\rangle_{\rm max}|$
also increases.
Figure~\ref{fig:7}(c) shows the maximum value
of $\langle V^{B}_x\rangle$ in the disordered regime,
$\langle V^{B}_x\rangle_{\rm max}$, versus $A$,
indicating that the drag effect on species B
increases with increasing ac drive amplitude.
The maximum value of
$\langle V^{B}_y\rangle$ in the disordered regime,
$\langle V^B_y\rangle_{\rm max}$, shows a similar increase as a function
of $A$ in Fig.~\ref{fig:7}(d).

\begin{figure}
\centering
\includegraphics[width=3.5in]{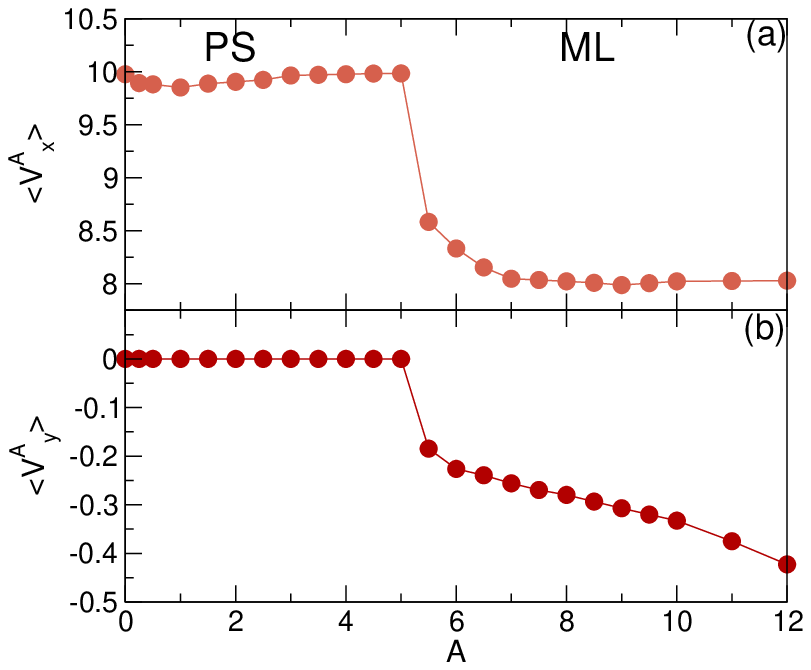}
\caption {(a) $\langle V^A_{x}\rangle$ vs $A$ at $F_{D} = 10.0$
for the system from Fig.~\ref{fig:4} with
$\rho=0.29$, $T/T_m=0.2$, and $\omega=0.02$.
The phase separated lane state (PS) 
and
moving liquid (ML) phase 
are highlighted.
(b) The corresponding $\langle V^A_{y}\rangle$ vs $A$ at
$F_D=10.0$. Near $A = 5.0$ the
system transitions from a phase separated lane state to a moving fluid.
}
\label{fig:8}
\end{figure}

To show more clearly how increasing the ac amplitude can
effectively melt the phase separated lane state,
in Fig.~\ref{fig:8}(a) we plot
$\langle V^A_{x}\rangle$ versus $A$ at $F_{D} = 10.0$
for the system from Fig.~\ref{fig:4}.
In this system, when $F_{D} < 5.0$,
$\langle V^A_{x}\rangle/F_{D}$ is close to $1.0$
since in the phase separated lane state
there are almost no collisions
between the driven and the chiral particles, and therefore the drag
is small.
For $A > 5.0$, the chiral orbits become large enough to place the
system in the disordered state
and the drag steadily increases as $A$ further increases,
leading to the drop in $\langle V^A_{x}\rangle$.
In Fig.~\ref{fig:8}(b) we show the corresponding value of
$\langle V^A_{y}\rangle$ versus $A$ at $F_{D} = 10.0$.
This quantity is close to zero in the phase separated lane state
and increases in magnitude in the disordered state.

\begin{figure}
\centering
\includegraphics[width=3.5in]{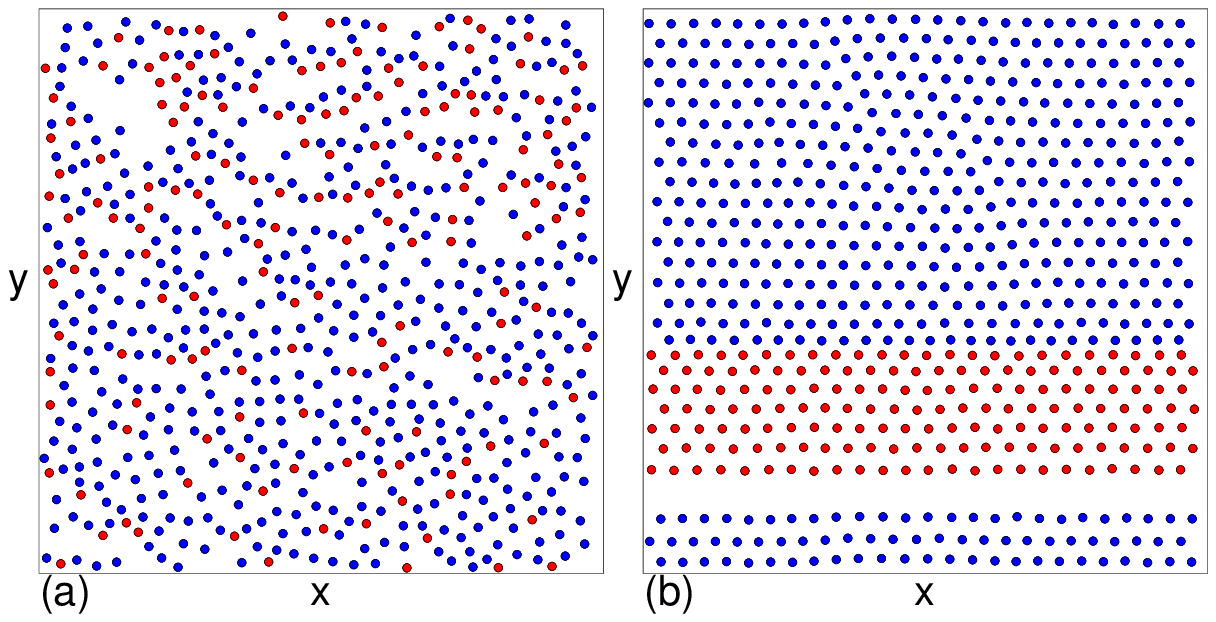}
\caption {The particle configurations at $A=8.0$ for the system from
Fig.~\ref{fig:7} with
$\rho=0.29$, $T/T_m=0.2$, and
$\omega=0.02$.
(a) A disordered state at $F_{D} = 2.0$. (b)
A phase separated lane state at $F_{D} = 20$.
}
\label{fig:9}
\end{figure}

In Fig.~\ref{fig:9}(a) we show the particle configurations
in the disordered regime for the system from
Fig.~\ref{fig:7} at an ac drive amplitude of $A=8.0$ and a dc drive of
$F_D=2.0$. Here there is a strongly mixed state in which both species A and
B have a finite velocity transverse to the dc driving direction.
For the same system at $F_D=20$,
Fig.~\ref{fig:9}(b) shows that there is a completely phase separated
lane state.
In the image, there is a spatial gap at one of the species interfaces, and
the species A region is compressed.
This is a result of the large orbit of species B, and occurs when,
during portions of the ac driving cycle, species B is moving predominantly
in the positive or negative $y$ direction and therefore exerts a net force on
the region filled with species A, opening a gap.
Due to the mutual repulsion between all
particles, when the ac driving cycle is aligned in the positive or negative
$x$ direction, species A relaxes back to form a uniformly dense system,
closing the gap.
In addition to formation of a gap, whenever the ac drive is oriented
along the $\pm y$ direction,
a compression of the lanes can occur. This compression
is reduced for smaller ac drive amplitudes.
In general, at the higher ac drive amplitudes,
the phase separated lane states are more ordered or show more
complete phase separation, while for lower
values of $A$, some driven particles
can become trapped in regions of active chiral particles.

\begin{figure}
\centering
\includegraphics[width=3.5in]{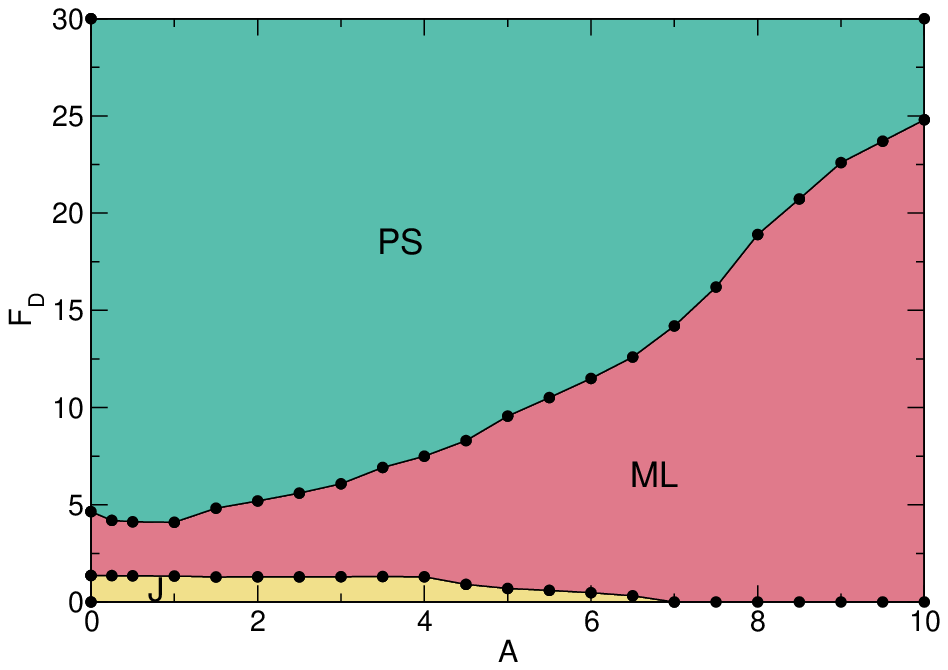}
\caption {Dynamic phase diagram as a function of $F_D$ vs $A$
for the system in Fig.~\ref{fig:2} with
$\rho=0.29$,
$T/T_m=0.2$, and $\omega=0.02$, showing
the jammed state (J, yellow), the moving liquid state (ML, red),
and the phase separated lane state (PS, green).
At large $A$ the jammed phase disappears and the transition between
the ML and PS states shifts to higher $F_D$.}
\label{fig:10}
\end{figure}

From the features in the transport curves,
we can construct the dynamical phase diagram shown in
Fig.~\ref{fig:10} as a function of $F_{D}$ versus ac
drive amplitude $A$, where we highlight the jammed phase,
the disordered moving liquid phase, and the phase separated lane state.
The jammed phase disappears for ac amplitudes greater than
$A=6.0$, where the system effectively melts even when $F_{D} = 0.0$.
The drive needed to induce the formation of
a phase separated lane state increases with increasing
ac drive amplitude.
There is no Hall effect within the jammed phase,
while in the disordered moving liquid state both species
exhibit a Hall effect.
If we use a random rather than periodic
initial placement of the species A particles,
the extent of the jammed phase is reduced but the
transition from the disordered moving liquid
to the phase separated lane state is unaffected.

The phase separated lane state we observe is similar to the
lane state found in binary driven Yukawa systems
where there are two species driven in opposite directions
\cite{Dzubiella02a,Glanz12}.
Similar laning has been seen in other systems of
oppositely driven particles \cite{Bain17,Isele23,Yu24}.
In these systems it is also possible to
obtain a low-drive disordered or fluid state
as well as ordered and disordered lane states at higher drives.
Other work showed that velocity-force curves in these types of systems
can be marked by signatures
of the transitions from jammed to disordered states
and from disordered to lane states \cite{Reichhardt18}.
The difference between the lane-forming systems studied previously and
the present work 
is the chiral nature of our species B particles,
which provides additional parameters that can be used
to control when the laning phases occur,
and also induces transverse or Hall velocities in the two different species.
There has been some work on laning pedestrian flows where the pedestrians
preferentially move to one side of the oppositely
walking pedestrians,
which has similarities to the chiral interactions \cite{Bacik23}.
In the case of the pedestrians,
the lanes become tilted, while in our system,
the lanes remain oriented with the driving direction.

\section{Thermal Effects}

\begin{figure}
\centering
\includegraphics[width=3.5in]{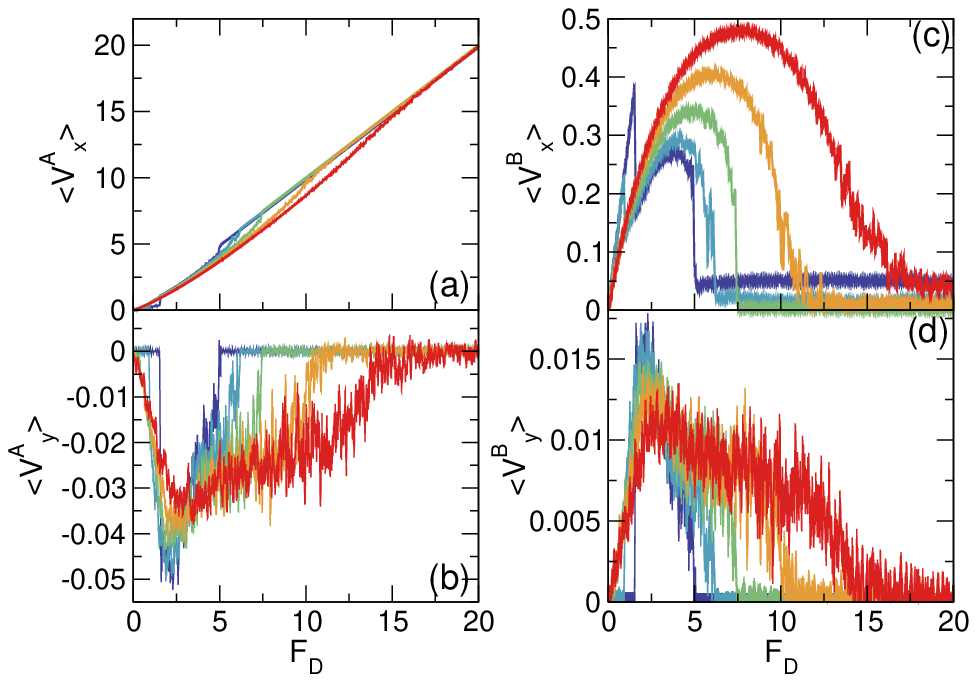}
\caption {Velocity curves for the system from Fig.~\ref{fig:2} with
$\rho=0.29$, $A=2.0$, and $\omega=0.02$ at temperatures
of $T/T_m=0.0$ (dark blue), 0.4 (light blue), 0.8 (green), 1.2 (yellow),
and 1.6 (red).
(a) $\langle V^A_{x}\rangle$ vs $F_{D}$.
(b) $\langle V^A_{y}\rangle$ vs $F_{D}$.
(c) $\langle V^B_{x}\rangle$ vs $F_{D}$.
(d) $\langle V^B_{y}\rangle$ vs $F_{D}$.
}
\label{fig:11}
\end{figure}

We next study how the dynamic phases evolve
under increasing thermal fluctuations.
For $F_{D} = 0.0$ and an ac amplitude of $A=0.0$,
the static system melts at a well-defined temperature $T_m$ that is
a function of the particle density.
In Fig.~\ref{fig:11}(a,b) we plot
$\langle V^A_{x}\rangle$ and
$\langle V^A_{y}\rangle$, respectively,
versus $F_D$ for the system from Fig.~\ref{fig:2} at temperatures of
$T/T_m = 0.0$, 0.4, 0.8, 1.2, and $1.6$.
The extent of the jammed phase
decreases with increasing $T/T_m$,
while the transition to the phase separated lane state shifts
to higher values of $F_{D}$.
The maximum magnitude of $\langle V^A_{y}\rangle$ decreases for higher $T/T_m$,
indicating that thermal fluctuations are reducing the Hall effect.
In Fig.~\ref{fig:11}(c,d) we plot the corresponding
$\langle V^B_{x}\rangle$ and $\langle V^B_{y}\rangle$,
respectively, versus $F_{D}$.
Here the maximum value of $\langle V^B_{x}\rangle$
increases with increasing temperature
while that of $\langle V^B_{y}\rangle$ decreases.
This indicates that although the Hall effect is reduced
by the increasing thermal fluctuations,
the drag effects are increased.

\begin{figure}
\centering
\includegraphics[width=3.5in]{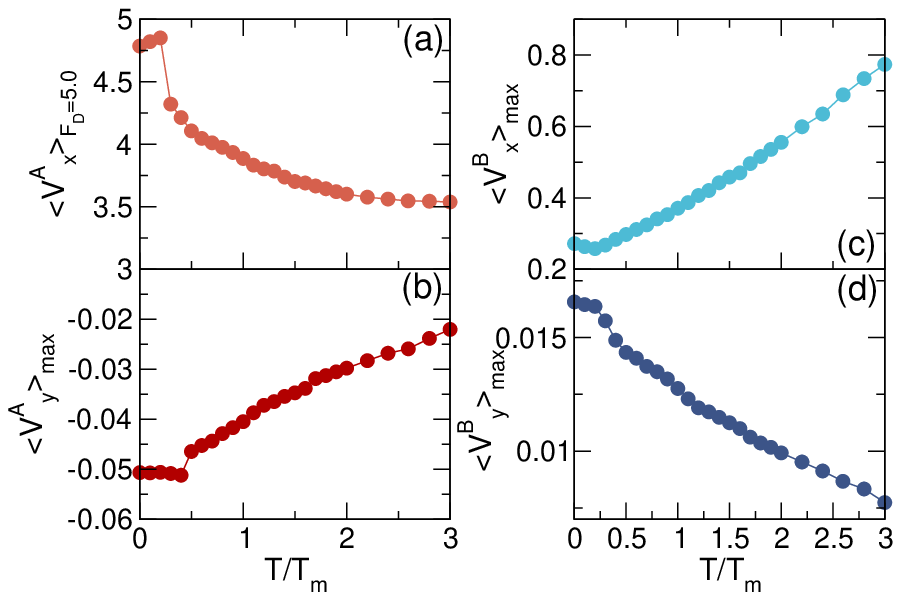}
\caption {Plots for the system from Fig.~\ref{fig:11} with
$\rho=0.29$, $A=2.0$, and $\omega=0.02$.
(a) The value of $\langle V^A_{x}\rangle$ at $F_D=5.0$,
$\langle V^A_x\rangle_{F_D=5.0}$, vs
$T/T_m$, showing a drop across the transition from
phase separated lane to moving fluid state.
(b) The maximum value of $\langle V^A_{y}\rangle$ in the disordered regime,
$\langle V^A_y\rangle_{\rm max}$, vs $T/T_m$.
(c) The maximum value of $\langle V^B_{x}\rangle$ in the disordered regime,
$\langle V^B_x\rangle_{\rm max}$, vs $T/T_m$.
(d) The maximum value of $\langle V^B_{y}\rangle$ in the disordered regime,
$\langle V^B_y\rangle_{\rm max}$, vs $T/T_m$.}
\label{fig:12}
\end{figure}

In Fig.~\ref{fig:12}(a,b) we plot the value of
$\langle V^A_{x}\rangle$ at $F_D=5.0$,
$\langle V^A_x\rangle_{F_D=5.0}$, and the maximum
value of $\langle V^A_{y}\rangle$ in the disordered regime,
$\langle V^A_y\rangle_{\rm max}$,  
versus $T/T_m$ for the system from Fig.~\ref{fig:11}.
For $T/T_m < 0.2$ the system is in a phase separated lane state and
$\langle V^A_{x}\rangle_{F_D=5.0}$ is close to constant,
while for $T/T_m > 0.2$ there is a disordered state and
$\langle V^A_{x}\rangle_{F_D=5.0}$
decreases with increasing $T/T_m$,
indicating an increase in the drag with increasing thermal fluctuations.
The higher temperature results in
increased mixing of the two species,
which leads to higher drag effects on the nondriven particles.
This is visible as an increase in the species B velocity
but a decrease in the velocity of species A.
The absolute value of $\langle V^A_{y}\rangle_{\rm max}$
drops with increasing $T/T_m$,
indicating a reduction in the transverse velocity with
higher temperature.
In Fig.~\ref{fig:12}(c) we plot the maximum value of
$\langle V^B_{x}\rangle$ in the disordered phase,
$\langle V^B_x\rangle_{\rm max}$, versus $T/T_m$, where we see
an increase with increasing thermal fluctuations.
This is consistent with
the simultaneous decrease in $\langle V^A_{x}\rangle_{F_D=5.0}$.
Figure~\ref{fig:12}(d) shows
the maximum value of $\langle V^B_{y}\rangle$ in the disordered phase,
$\langle V^B_y\rangle_{\rm max}$,
versus $T/T_m$.
We find that
$\langle V^B_y\rangle_{\rm max}$
decreases with increasing $T/T_m$,
indicating that the Hall effect on species B is decreasing
as the temperature increases.

\begin{figure}
\centering
\includegraphics[width=3.5in]{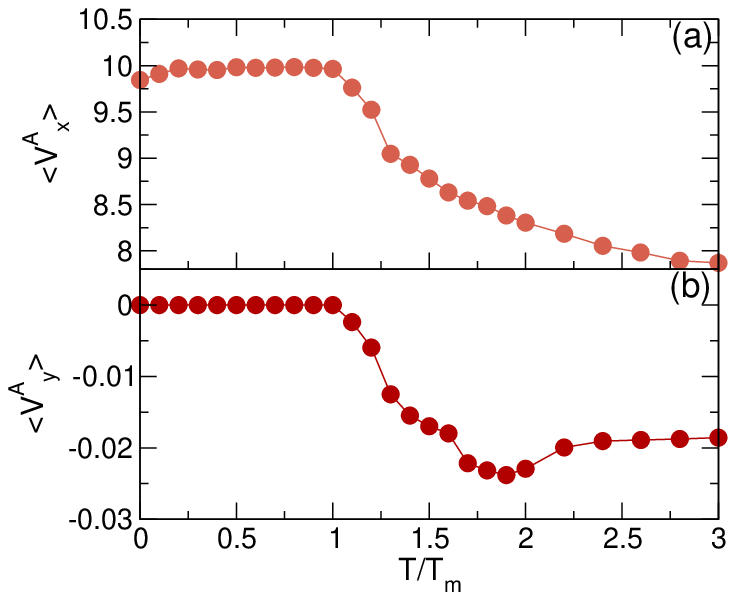}
\caption {Species A velocities for the system from Fig.~\ref{fig:11}
with
$\rho=0.29$, $A=2.0$, $\omega=0.02$, and
$F_D=10.0$.  
(a) $\langle V^A_{x}\rangle$ vs $T/T_m$.
(b) $\langle V^A_{y}\rangle$ vs $T/T_m$.
}
\label{fig:13}
\end{figure}

In Fig.~\ref{fig:13}(a,b) we plot
$\langle V^A_{x}\rangle$ and $\langle V^A_y\rangle$, respectively,
at $F_D=10.0$ versus $T/T_m$. We find a melting-like transition near
$T/T_m=1.0$, which coincides with the onset of a finite $\langle V^A_y\rangle$
or Hall effect.
At higher values of $F_{D}$ such as this, the
system is much more crystalline, and therefore
the transition to disordered flow occurs
close to the same temperature at which the
$F_{D} = 0.0$ system melts.

\begin{figure}
\centering
\includegraphics[width=3.5in]{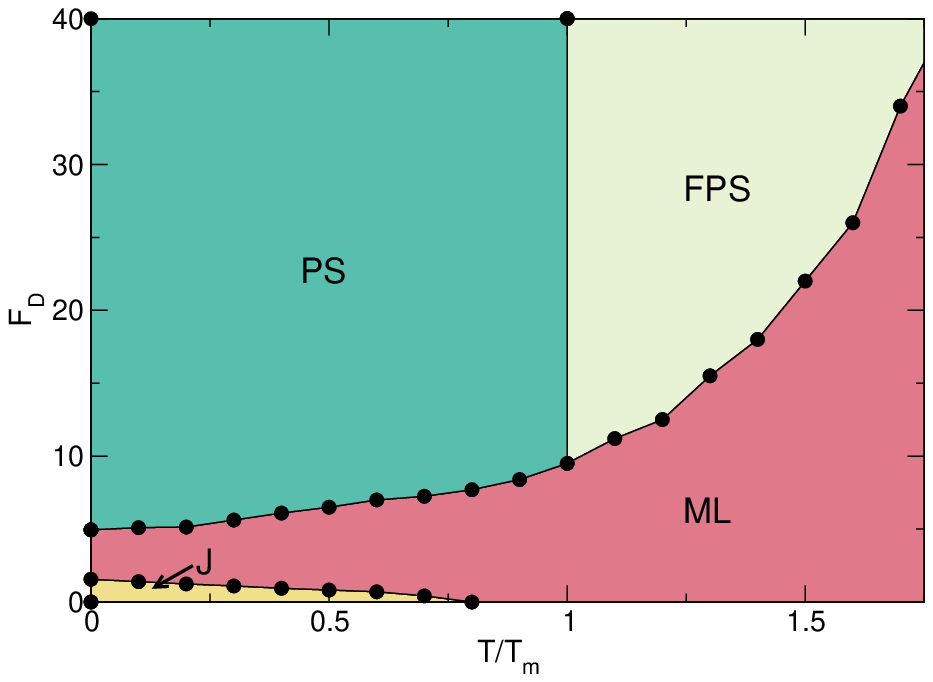}
\caption {Dynamic phase diagram as a function of $F_D$ vs $T/T_m$ for the
system from Fig.~\ref{fig:11} with
$\rho=0.29$, $A=2.0$,
and $\omega=0.02$, showing
the jammed state (J, yellow), the moving liquid state (ML, red),
the crystalline phase separated lane state (PS, dark green),
and the fluid phase separated lane state (FPS, light green).}
\label{fig:14}
\end{figure}

\begin{figure}
\centering
\includegraphics[width=3.5in]{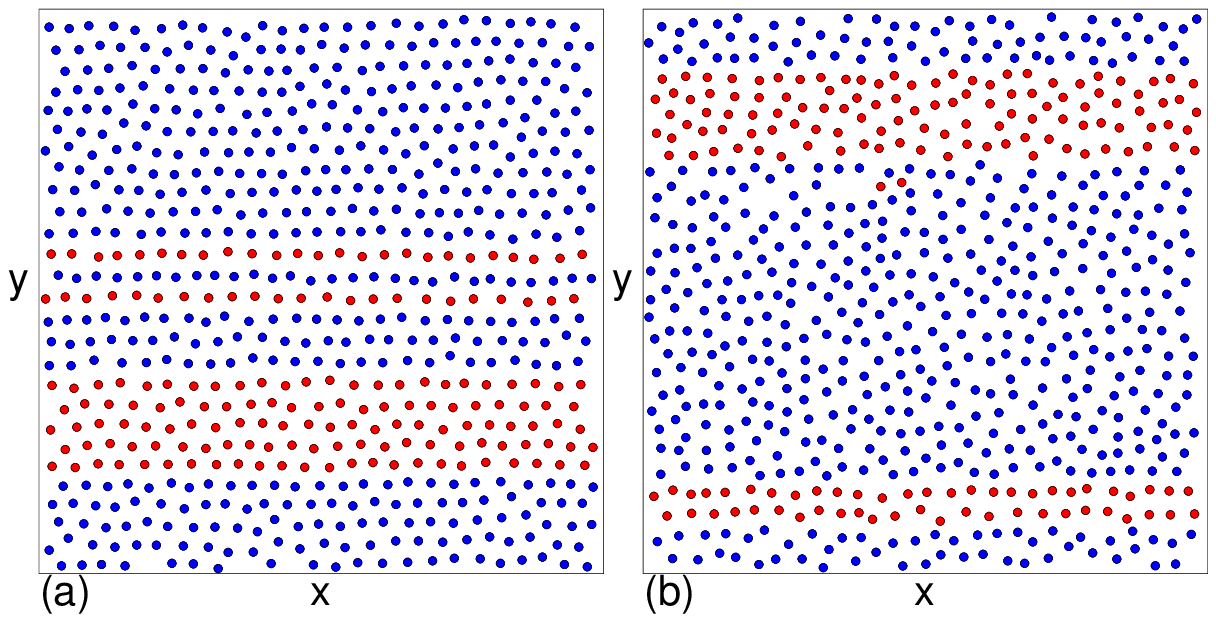}
\caption {The particle configurations for the system from Fig.~\ref{fig:14}
with
$\rho=0.29$, $A=2.0$ and $\omega=0.02$.
(a) The crystalline phase separated lane state at
$F_{D} = 10.0$ and $T/T_m = 0.8$.
(b) The fluid phase separated lane state at
$F_{D} = 15$  and $T/T_m = 1.2$.
}
\label{fig:15}
\end{figure}

From the velocity-force curves and images, we construct a
dynamic phase diagram as a function of $F_{D}$ versus $T/T_m$, shown in
Fig.~\ref{fig:14}, for the system in Figs.~\ref{fig:11} and \ref{fig:12}.
The jammed phase disappears with increasing temperature,
and the drive at which the phase separated lane states appear
increases with
increasing $T/T_m$.
We find that the moving crystalline phase
separated lane state that appears for $T/T_m < 1.0$
changes to a fluid phase separated lane state for $T/T_m > 1.0$.
In Fig.~\ref{fig:15}(a) we illustrate the crystalline phase separated
lane state
at $F_D=10.0$ and $T/T_m = 0.8$,
while in Fig.~\ref{fig:15}(b) we show that at
$F_D=15$ and $T/T_m = 1.2$,
the system is disordered but still contains well defined lanes of each
particle species.

In general, the dynamic phase diagram in
Fig.~\ref{fig:14} has similarities to the dynamic phase diagram
obtained for driven particles moving over
random quenched disorder \cite{Reichhardt17}.
Systems such as these, which include vortices in type-II superconductors,
form a crystal in the absence of quenched disorder and drive,
and have a well-defined melting temperature $T/T_m$
that is characterized by the proliferation of topological defects.
When strong quenched disorder is present, there is
a disordered pinned state at low drives,
a disordered plastic flow state at intermediate drives,
and a moving crystal or smectic state at high drives.
The drive needed to reach the ordered state increases
with increasing $T/T_m$ and diverges at the equilibrium melting
temperature.
In our system, there is no quenched disorder,
so it is still possible for a crystalline state to form
at finite drives up to $T/T_m$,
while for temperatures above $T/T_m$,
the system 
can form a phase separated lane state
when the drive is strong enough,
meaning that the system is not a completely random
fluid. 

\section{Changing Frequency}

\begin{figure}
  \centering
\includegraphics[width=3.5in]{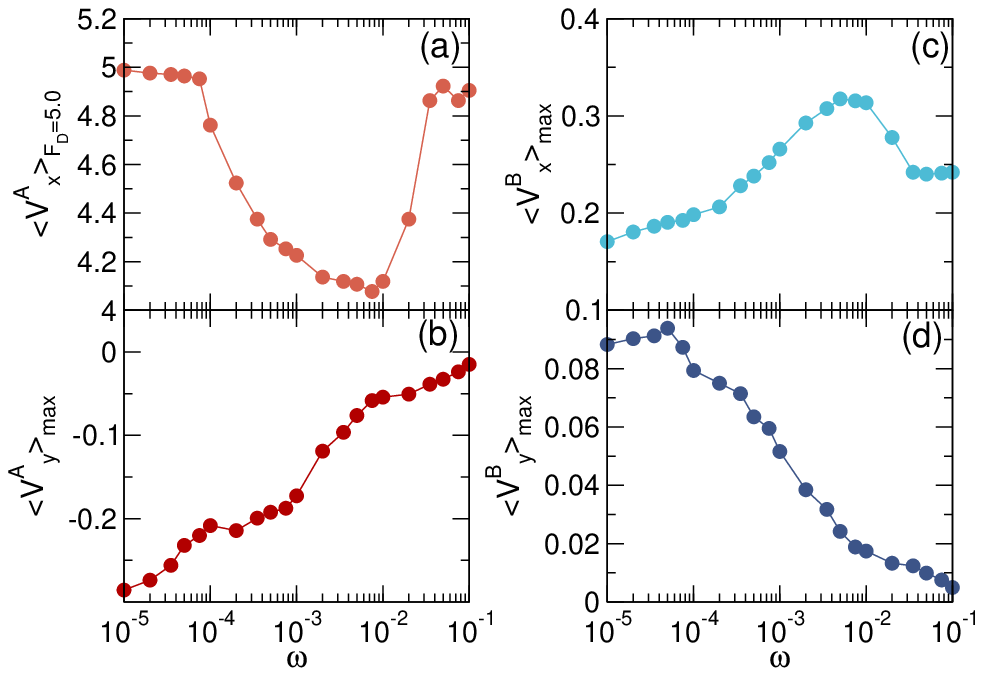}
\caption {Plots for a system with
$\rho=0.29$,
$A=2.0$, and $T/T_m=0.2$.
(a) The value of $\langle V^A_{x}\rangle$ at $F_D=5.0$,
$\langle V^A_x\rangle_{F_D=5.0}$, vs
$\omega$, showing a drop across the transition from
phase separated lane to moving fluid state.
(b) The maximum value of $\langle V^A_{y}\rangle$ in the disordered regime,
$\langle V^A_y\rangle_{\rm max}$, vs $\omega$.
(c) The maximum value of $\langle V^B_{x}\rangle$ in the disordered regime,
$\langle V^B_x\rangle_{\rm max}$, vs $\omega$.
(d) The maximum value of $\langle V^B_{y}\rangle$ in the disordered regime,
$\langle V^B_y\rangle_{\rm max}$, vs $\omega$.}
\label{fig:16}
\end{figure}

The size of the chiral orbits can be altered by varying
the ac drive amplitude $A$ and/or the drive frequency $\omega$.
These two control parameters
have some differences
because the ac drive amplitude changes the net circular force,
whereas the ac drive frequency only changes the orbit size.
In Fig.~\ref{fig:16}(a) we
plot the value of $\langle V^A_{x}\rangle$ at $F_D=5.0$,
$\langle V^A_x\rangle_{F_D=5.0}$,
versus $\omega$ for an ac amplitude of $A=2.0$.
At this drive, the system is in the phase separated state
for $\omega > 0.1$ and in the
disordered flow state for $0.0001 < \omega \leq 0.1$.
The transition between these two states is marked
by a drop in $\langle V^A_{x}\rangle_{F_D=5.0}$
at $\omega=0.1$ as the frequency decreases.
There is a local minimum in $\langle V^A_x\rangle_{F_D=5.0}$
near $\omega = 0.01$.
In the phase separated regime,
$\langle V^A_{x}\rangle_{F_D=5.0}$ is close to $5.0$,
indicating that the species A particles are moving in
nearly free flow.
In Fig.~\ref{fig:16}(b) we plot the maximum value of
$\langle V^A_{y}\rangle$ in the disordered regime,
$\langle V^A_y\rangle_{\rm max}$, versus $\omega$.
As $\omega$ increases,
the magnitude of $\langle V^A_{y}\rangle_{\rm max}$ decreases, indicating
that the maximum Hall angle is also becoming smaller.
This reduction in the Hall response results because
as $\omega$ increases, the chiral orbit becomes smaller,
reducing the importance of chiral effects.
For smaller $\omega$, the chiral orbits are bigger
and the Hall effect increases in size.
In Fig.~\ref{fig:16}(c) we plot the maximum value
of $\langle V^B_{x}\rangle$ in the disordered regime,
$\langle V^B_x\rangle_{\rm max}$,
versus $\omega$.
This quantity passes through a maximum near
$\omega = 0.01$, which also corresponds to the local minimum in
$\langle V^A_{x}\rangle_{F_D=5.0}$, as shown in Fig.~\ref{fig:16}(a).
In Fig.~\ref{fig:16}(d) we plot 
the maximum value of $\langle V^B_{y}\rangle$ in the disordered regime,
$\langle V^B_y\rangle_{\rm max}$,
versus $\omega$. Here we find
an increase with decreasing $\omega$,
indicating an increased Hall response of species B as the orbit becomes
larger.

\begin{figure}
\centering
\includegraphics[width=3.5in]{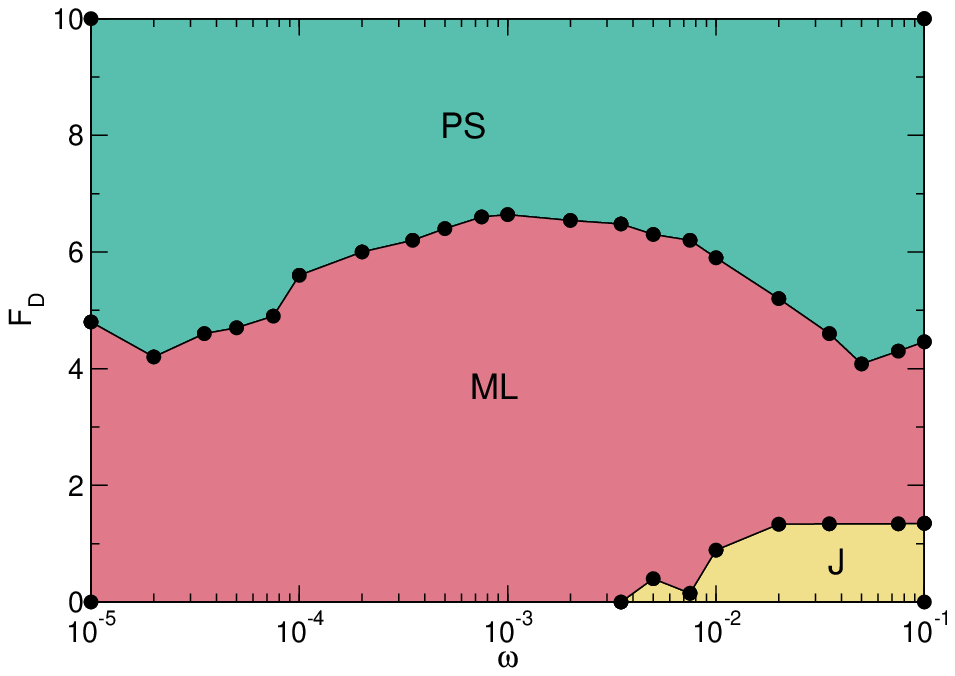}
\caption {Dynamic phase diagram as a function of $F_D$ vs $\omega$ for
the system from Fig.~\ref{fig:16} with
$\rho=0.29$,
$A=2.0$, and $T/T_m=0.2$, showing the
jammed state (J, yellow), the moving liquid state (ML, red),
and the crystalline phase separated lane state (PS, green).}
\label{fig:17}
\end{figure}

We use the features in the transport curves
to construct a dynamic phase diagram,
shown in Fig.~\ref{fig:17}, as a function of $F_{D}$ versus $\omega$, where
we highlight the jammed, disordered moving liquid, and crystalline phase
separated lane states.
The jammed phase disappears for $\omega < 0.0035$,
and the maximum extent of the disordered
moving liquid phase occurs near $\omega = 0.001$.
For lower $\omega$, even though the orbits of the chiral
particles are large, the system is still able to form
a phase separated lane state at sufficiently large drive $F_D$.

\begin{figure}
\centering
\includegraphics[width=3.5in]{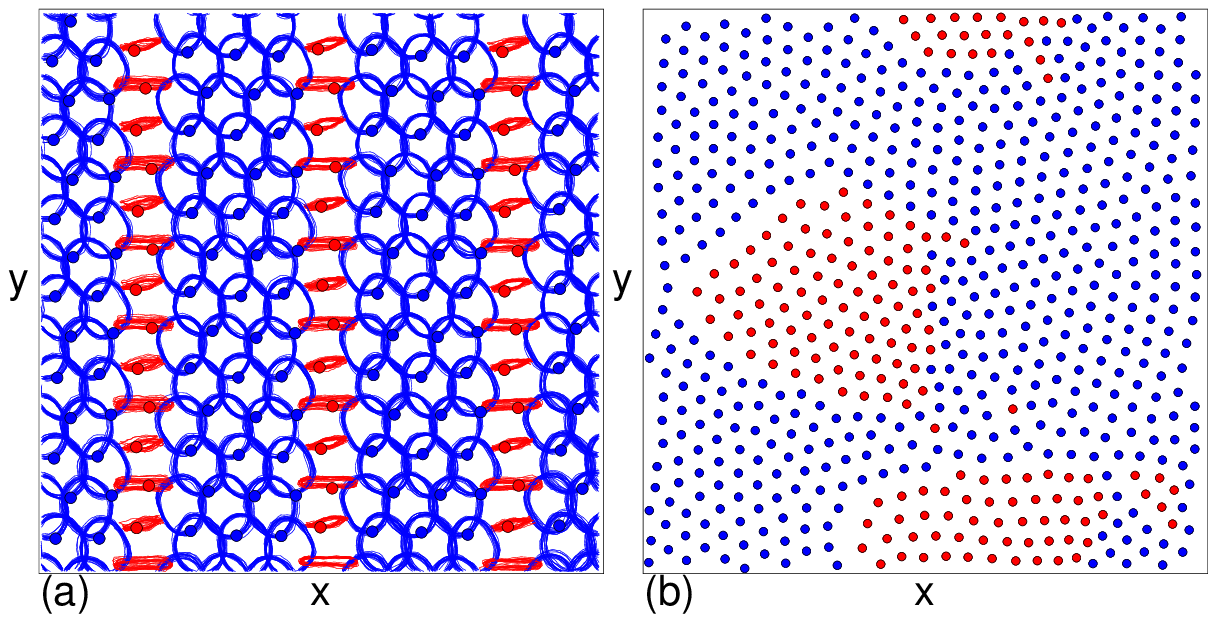}
\caption {Images for the system from Fig.~\ref{fig:17} with
$\rho=0.29$, $A=2.0$, $T/T_m=0.2$, and $\omega=0.0075$.  
(a) Particle configurations and trajectories at $F_{D} = 0.0$.
(b) Particle positions only at $F_{D} = 0.1$ in the
phase separated state.}
\label{fig:18}
\end{figure}

Another interesting feature in Fig.~\ref{fig:17} is that
near $\omega = 0.0075$, the jammed phase is strongly suppressed,
but there is a reappearance of the jammed state
for $\omega = 0.005$.
This occurs since, at the $\omega=0.005$ orbit size,
the chiral orbits just barely overlap
and the system becomes phase separated even close to $F_{D} = 0.0$.
In Fig.~\ref{fig:18}(a), we show the particle
positions and trajectories at $F_{D} = 0.0$ for
$\omega = 0.0075$, where the overlapping chiral orbits of species B
are clearly visible.
In addition, species A shows strong quasi-one-dimensional correlated
motion due to the coupling with the chirally moving particles.
Once even a small dc drive is applied,
the system destabilizes and breaks
up into a phase separated state such as
the one shown in Fig.~\ref{fig:18}(b) at $F_{D} = 0.2$.
This phase separated state is distinct from the
high drive phase separated lane state.
There are no lanes present; instead, patches of species A appear
inside a background of species B.
These patches or clusters exhibit a large scale rotation and
have a finite Hall effect,
whereas the laned states do not rotate and the Hall effect is absent.
At higher drives, the phase separated state
illustrated in Fig.~\ref{fig:18}(b) breaks into a
disordered flow state before forming a phase separated lane state
at higher drives.
There can be additional interesting effects
produced by commensurate orbits for different values
of $\omega$ at $F_{D} = 0.0$, which will be studied elsewhere. 

\section{Changing Density and Discussion}

\begin{figure}
\centering
\includegraphics[width=3.5in]{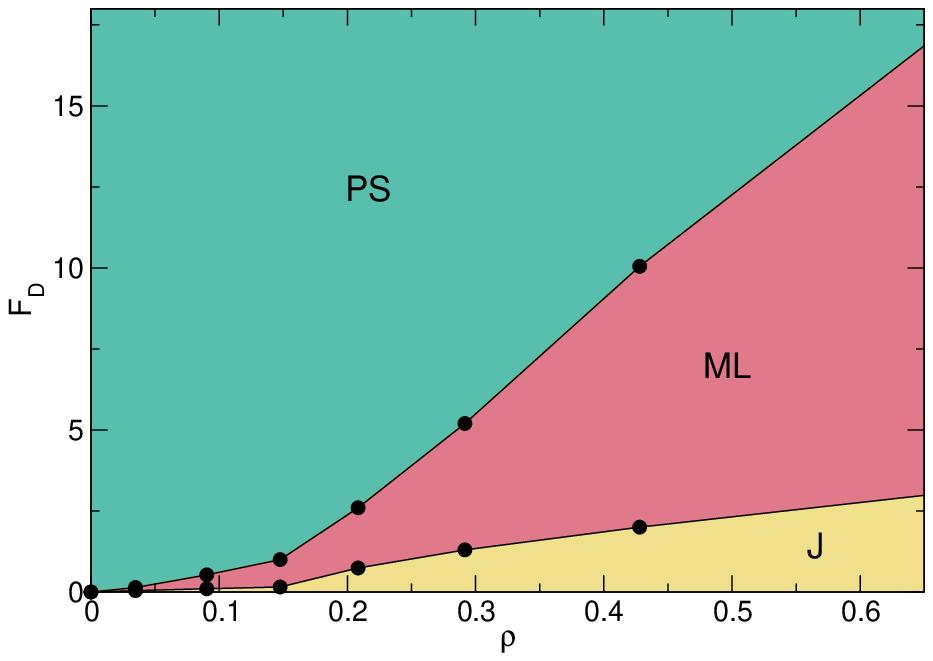}
\caption {Dynamic phase diagram as a function of $F_{D}$ vs $\rho$
for a system with
$A=2.0$, $\omega=0.02$, and $T/T_m=0.2$,
showing the jammed state (J, yellow), the moving liquid state (ML, red),
and the crystalline phase separated lane state (PS, green).}
\label{fig:19}
\end{figure}

To explore the effects of changing the particle density.
in Fig.~\ref{fig:19} we plot a
dynamic phase diagram as a function of $F_{D}$ versus $\rho$ for
the same system from Fig.~\ref{fig:2}
with an ac drive amplitude of $A=2.0$ at $\omega = 0.02$.
The jammed phase grows in extent with increasing $\rho$,
while the drive at which the
transition from the disordered to the phase separated lane state
occurs increases.
The $F_D$ versus $\rho$ phase diagram is similar to the
$F_D$ versus $A$ phase diagram from Fig.~\ref{fig:10},
but for increased density the extent of the jammed phase grows,
while for increased ac drive amplitude it decreases.

We varied the value of $N_d/N$ over the range of $0.05$ to $0.5$, and
found results very similar to what is shown here for $N_d/N=0.25$.
We also considered the case of particles with Bessel function interactions
and find very similar behavior as the Yukawa system, indicating
that our results should be general to any type of
driven particle system moving through a chiral bath of other particles.
In this work we focused on systems with intermediate-range interactions.
One could also consider the effects of introducing
competing long-range repulsion and short-range attraction,
or even what happens when there are three particle species
instead of two.
Another possibility would be to fix
the rotating particles to orbit around specific locations,
so that phase separation is not allowed to occur.
In that case, the transverse response could be enhanced.
We have only studied this system under a single ramp up of the
dc drive. If the dc drive were subsequently reduced back to zero,
hysteresis or memory effects could appear.
This would be most likely to occur for smaller ac amplitudes
where the lane state could persist to smaller drives on the
ramp down compared to the drive at which it first appears during the
ramp up.
For larger ac amplitudes, the system will fluidize
even at zero drive, and the hysteresis could be lost, with the appearance
and disappearance of the phase separated lane state occurring at the same
value of $F_D$ for both the ramp up and the ramp down.

\section{Summary}

We have examined the dynamics of collectively
interacting Yukawa particles driven through an
array of other Yukawa particles that are undergoing
circular motion but do not couple to the external drive.
At low drives, the system forms a jammed elastic solid where
the driven particles are able to drag the chiral particles.
At higher drives the driven particles break away from the
chiral particles, leading to the appearance of plastic deformations and
formation of a disordered state.
Due to scattering from the chiral particles,
in the disordered state the driven particles have a finite
transverse velocity or Hall effect.
At even higher drives, the system phase separates into a
lane state and the Hall effect vanishes because
the frequency of collisions between the driven and chiral particles
is strongly reduced.
In the disordered regime, the driven particles
induce a strong drag effect on the chiral particles,
giving the chiral particles a finite velocity in the direction of
the dc drive. At the same time, the dragged chiral particles
exhibit a Hall effect in the direction opposite to that of
the driven particles.
As the ac driving amplitude on the chiral particles is increased,
the extent of the jammed phase is reduced,
and the drive at which the transition between the disordered and
phase separated lane state occurs increases.
The drag response and transverse response both increase with increasing
ac drive amplitude.
If the ac drive amplitude is held fixed but the thermal fluctuations are
increased,
the drag effect exerted by the driven particles on the chiral particles
increases, but the transverse response or Hall effect of both species
decreases.
The phase separated lane state is crystalline in nature
at low temperatures, but becomes fluidlike
at high temperatures.
For increasing ac frequency at
fixed ac drive amplitude, the chiral orbits become smaller, reducing the
importance of the chirality and decreasing the transverse response of both
species.
Our results show that when driven particles move through a chiral
background of undriven particles, a transverse velocity or Hall effect
appears for both the driven species and the chiral background particles.
These results should be general for a wide class of systems.

\ack{We gratefully acknowledge the support of the U.S. Department of
Energy through the LANL/LDRD program for this work.
This work was supported by the US Department of Energy through
the Los Alamos National Laboratory.  Los Alamos National Laboratory is
operated by Triad National Security, LLC, for the National Nuclear Security
Administration of the U. S. Department of Energy (Contract No. 892333218NCA000001).}

\section*{References}
\bibliographystyle{iopart-num}
\bibliography{mybib}

\end{document}